# LED receiver impedance and its effects on LED-LED visible light communications

Shangbin Li,[1,†] Boyang Huang,[1] and Zhengyuan Xu[1,2*]
[1] *Key Laboratory of Wireless-Optical Communications, Chinese Academy of Sciences, University of Science and Technology of China, Hefei, Anhui 230027, China.*
[2] *Shenzhen Graduate School, Tsinghua University, Shenzhen 518055, China.*
*xuzy@ustc.edu.cn; †shbli@ustc.edu.cn

**Abstract:** This paper experimentally demonstrates that the AC impedance spectrum of the LED as a photodetector heavily depends on the received optical power, which may cause the impedance mismatch between the LED and the post trans-impedance amplifier. The optical power dependent impedance of the LED is well fitted by a modified dispersive carrier transport model for inorganic semiconductors. The bandwidth of the LED-LED visible light communication link is further shown to decrease with the optical power received by the LED. This leads to a trade-off between link bandwidth and SNR, and consequently affects the choice of the proper dada modulation scheme.

**Introduction**

Visible light communication (VLC) undergoes rapid development. There are various advantages compared with traditional wireless communication, such as free spectrum license, resistance to electromagnetic interference and high directionality which can be used in secure communication and accurate positioning. Among different light sources, LEDs are energy efficient, and simultaneously provide high quality illumination and high data rate transmission.

Most researches are focused on the VLC using the photodiode (PD) or the avalanche photodiode (APD) at the receiver [1–3]. There are researches using LEDs as photon detectors at the receiver [4–14]. At the very beginning, some researchers noticed that an LED could be excited by photons. One of these excitations is photoluminescence (PL), i.e., the LED emits the light after absorption of photons [4,15]. This light to light conversion has been used to test the structure of LEDs. Accompanied by PL, its photoelectric (PE) responsibility implies that an LED can function as a photodiode. The LED is used as spectrally selective photometer for its narrow spectral width [5]. About 20 years ago, Miyazaki et al. used a Zn-doped InGaN blue LED and a GaAlAs red LED as photodetectors [6]. They investigated the emission and extinction spectra of blue and red LED as well as the impulse response and the influence of reverse bias. However, LEDs have been mainly used either as light sensors or as detectors for simple and low data rate systems for a decade [7-10]. Recently, the LED receiver has attracted much attention in visible light communication. Because it can be used both as the transmitter and receiver, it is convenient to construct a simple duplex VLC system, utilizing only two LEDs instead of two LEDs and two PDs [11]. Later, two research groups used selected commercial red LEDs and reached data rates more than 100Mbps with DMT or OFDM modulation [12,13]. Furthermore, Kowalczyk et al. investigated the influence of the reverse bias on LED-LED VLC links [14].

The bandwidth of the LED-LED link heavily depends on the AC impedance of the LED receiver. It is thus necessary to study the influence of the received optical power on the AC impedance of the LED receiver to better understand LED-LED VLC links. Our experimental results show the AC impedance spectrum of the LED shifts towards the lower frequency with the larger received optical power. The AC impedance spectroscopy of the LED receiver without trans-impedance amplifier (TIA) exhibits the phase transition-like behavior. At the critical frequency point, the AC impedance is pure resistance, and the critical frequency point shifts with the optical power. We slightly modify the dispersive carrier transport model of inorganic semiconductors, and quantitatively interpret the optical power-dependence of the LED impedance. Furthermore, we experimentally demonstrate the signal to noise ratio (SNR) and the bandwidth of the LED-LED link are reversely correlated. Increasing the optical power received by an LED receiver, the communication performance may be deteriorated. The bandwidth decreases with the optical power and causes more bit errors. For clarifying how to make the tradeoff between bandwidth and SNR to reach the highest data rate of the LED-LED visible light link, the frequency response curves are used to simulate the communication performances with three different modulations. We find different modulations have different sensitivities to bandwidth and SNR. OOK performs well in the low optical power region while higher order PAM works better in the higher optical power region.

## Effects of the injected optical power on the LED receiver impedance

Impedance spectroscopy is a powerful tool for investigating the intrinsic relaxation processes and device structures of organic and inorganic materials and devices, such as the frequency dependence of the conductivity, the dielectric constant, and dopant and trap concentrations as well as their spatial distributions [16]. It has been shown the impedance spectrum depends on the bias voltage and the junction temperature [17]. However, few researches have revealed the relations between the optical field and the impedance spectrum of the LEDs.

We measured the impedance of the orange-red LED by connecting the LED directly to a vector network analyzer without trans-impedance amplifier, and record the Smith chart. Using the reflection coefficients read from the Smith chart, we can calculate the impedance of LED at certain frequency by Eq. (1) below

$$Z = \frac{1-\Gamma}{1+\Gamma} Z_0 \qquad (1)$$

where $\Gamma$ is reflection coefficient and $Z_0 = 50\Omega$ is the output impedance of the vector network analyzer.

It is found that the AC impedance spectroscopy of the orange-red LED receiver without a trans-impedance amplifier exhibits the phase transition-like behavior as shown in Figure 1. At the transition point about 12MHz, the AC impedance is pure resistance at a value of near 1.5$\Omega$. Moreover, the AC impedance spectrum of the LED tends to shift towards the lower frequency with the larger received optical power. The critical frequency point also shifts with the optical power, and the corresponding critical impedance increases with the optical power.

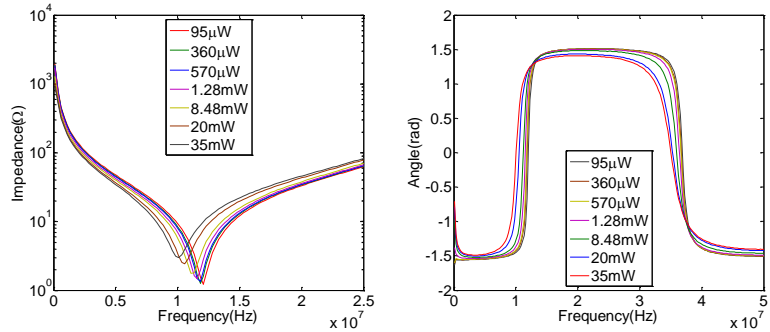

Fig. 1. The experimental AC impedance spectrum (left: amplitude, right: phase) of the orange-red LED receiver without the bias voltage added under different injected optical power.

Up to date, the intrinsic mechanism of the light-dependent impedance still lacks of a reasonable theoretical model. The influence of the injected optical signal on the AC impedance can be approximately equivalent to that of an effective reverse bias acting on the LED. According to the theoretical model of single-carrier transport for inorganic semiconductors, i.e., the current-flow and Poisson's equations, one can obtain the following analytical expression of the impedance under the nondispersive transport condition [17]

$$Z(\omega) = \frac{1}{\frac{1}{R(\omega)} + i\omega C(\omega)} \tag{2}$$

where

$$R(\omega) = \frac{6\left[(\Omega - \sin\Omega)^2 + \left(\frac{\Omega^2}{2} + \cos\Omega - 1\right)^2\right]}{g\Omega^3(\Omega - \sin\Omega)} \tag{3}$$

$$\omega C(\omega) = \frac{g\Omega^3\left(\frac{\Omega^2}{2} + \cos\Omega - 1\right)}{6\left[(\Omega - \sin\Omega)^2 + \left(\frac{\Omega^2}{2} + \cos\Omega - 1\right)^2\right]} \tag{4}$$

where $g$ is the steady current incremental conductance given by

$$g = \frac{3C_\lambda}{\tau} \tag{5}$$

and $\Omega = \omega\tau$ is the transit angle, where $C_\lambda$ and $\tau$ are the geometrical capacitance and the carrier transit time, respectively. Substituting Eqs. (3) and (4), Eq. (2) is simplified to

$$Z(\omega) = \frac{6\left[(\Omega - \sin\Omega) - i\left(\frac{\Omega^2}{2} + \cos\Omega - 1\right)\right]}{g\Omega^3} \tag{6}$$

Equation (6) can well fit the experimental AC impedance at low frequencies as shown in Figure 2. However, for a larger frequency, the nondispersive transport model seems to fail to interpret the experimental AC impedance of the LED receiver even if the localized states are taken into account [17]. Ref. [18] presents a dispersive transport model of the charge carrier in an organic LED, in which the mobility of the carrier depends on the frequency of the small signal. In this situation, the $sin\Omega$ and $cos\Omega$ terms in Eq. (6) become $sin[\Omega\mu(\Omega)]$ and $cos[\Omega\mu(\Omega)]$, respectively. $\mu(\Omega)$ represents the frequency-dependent mobility of the carrier, and is given by

$$\mu(\Omega) = \frac{4}{3\left[1 + M(i\Omega)^{1-\alpha}\right]} \tag{7}$$

where $M$ and $\alpha$ are the dispersive parameters. This frequency-dependent mobility of the carrier could bring new kind of nonlinearity of the LED impedance. Preliminary numerical results have shown that a slightly modified dispersive transport model of the AC impedance can interpret our experimental data. Those results are depicted in Figure 3, where the theoretical fitting curves are given by the following equations

$$Z_1(\omega) = Z(\omega) + \frac{a}{\cos^2(k\omega+\varphi_0)} + ib\cdot\tan(k\omega+\varphi_0) \tag{8}$$

The last two terms on the right side of Eq. (8) are added for approximately describing the effect of the frequency-dependent mobility on the impedance. From Figure 3, it is implied that the geometrical capacitance increases and the carrier transit time decreases with the injected optical power.

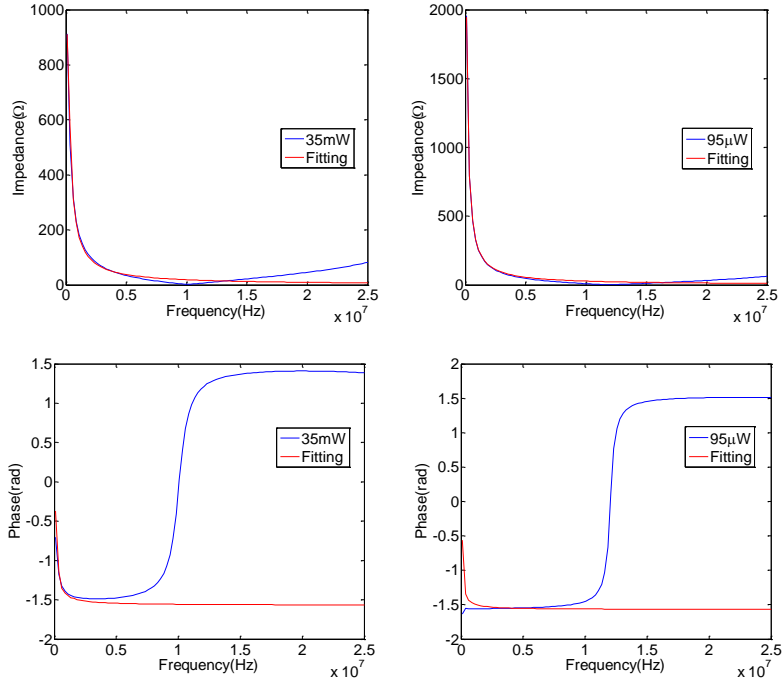

Fig. 2. The experimental and theoretical results of the AC impedance of the orange-red LED without the added bias voltage under two values of optical power.

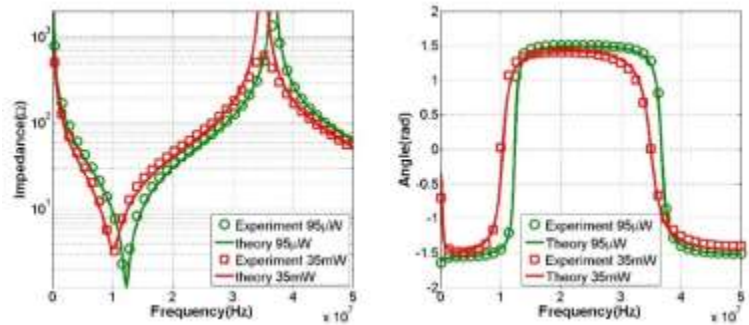

Fig. 3. The experimental and theoretical fitting results of the AC impedance of the orange-red LED without the added bias voltage under two values of optical power. The theoretical results are given by Eq. (8) with $\tau=350\mu s$, $C_\lambda=3.8nF$, $a=1.02\Omega$, $b=45.8\Omega$, $k=7.44\times10^{-9}s$, $\varphi_0=-0.1487$ for the case of 95μW injected light, and $\tau=11\mu s$, $C_\lambda=4.8nF$, $a=2.65\Omega$, $b=44.7\Omega$, $k=7.44\times10^{-9}s$, $\varphi_0=-0.0669$ for the case of 35mW injected light.

Furthermore, based on the optimal fitness of the experimental data and the model in Eq. (8) in the frequency range (0, 50MHz), the steady current incremental conductance and the carrier transit time versus the injected optical power are plotted in Figure 4. For $\Omega \gg 1$, the impedance described by Eq. (8) exhibits an oscillation with the period $1/k$. However, the actual measured impedance oscillates with varying period. Thus, in the larger frequency range, we should introduce the frequency-dependent perturbation in $k$. The related work will be discussed elsewhere.

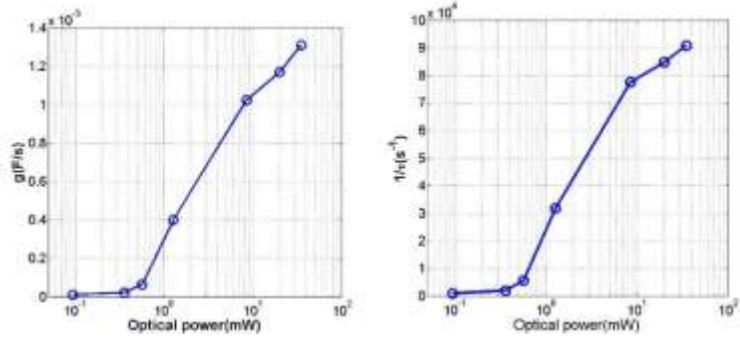

Fig. 4. The steady current incremental conductance and the carrier transit rate versus the injected optical power.

## Bandwidth dependence of the LED-LED link on the optical power

In this section, we investigate the frequency spectral response curves of the LED-LED links under different optical power. The testing system includes two parts. Figure 5 gives pictures of the transmitting part and receiving part. The transmitting part contains an orange-red LED with a converging lens as the transmitter. A frequency sweep signal is generated and added on the LED transmitter by an arbitrary waveform generator (AWG). The DC bias voltage applied to the LED is 1.89V and peak to peak voltage (Vpp) is 0.14V when frequency is 5MHz. The transmitting part does not change parameters during the experiment, while the optical power is changed by changing the distance between the transmitter and the receiver. In the receiving part, there is the same orange-red LED with a trans-impedance amplifier (TIA). The amplifier output is connected to a spectrum analyzer to measure the spectral response. An optical power meter is placed at the position of the LED receiver before each measurement. When the light spot is large enough and the distribution is uniform, the ratio of the optical power measured by the optical power meter and the optical power received by the LED remains a constant. Throughout this paper, we use the optical power detected by the optical power meter instead of the power received by the LED for convenience.

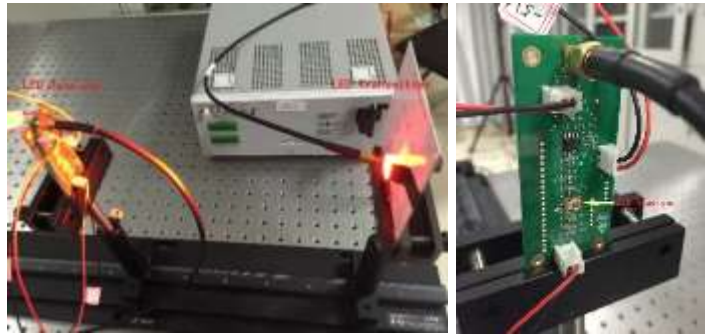

Fig. 5. Pictures of LED transmitter and LED receiver. The transmitter (left) is an LumiLED Rebel series orange-red LED with a converging lens, and the receiver (right) is another LumiLED Rebel orange-red LED with a trans-impedance amplifier.

The frequency response curves of the LED-LED link are shown in Figure 6. It is clearly observed that with the reduction of optical power, the total response intensity decreases but the response curve decays slowly, thus the higher bandwidth can be reached. The −3dB, −10dB, and −20dB bandwidths versus optical power are plotted in Figure 7. Orders of MHz to tens of MHz bandwidths can be achieved. Similar results can be obtained when using a royal-blue LED as the transmitter and a green LED as the receiver.

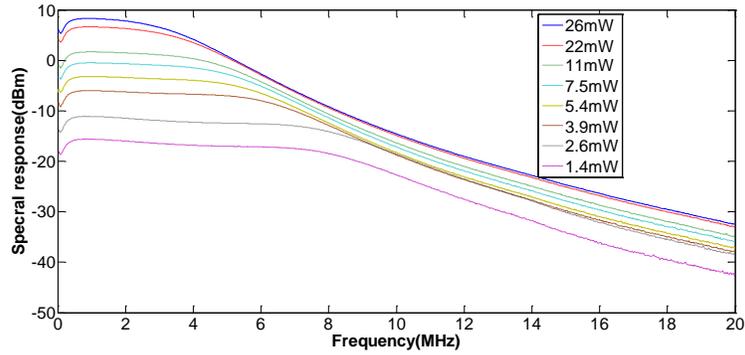

Fig. 6. Frequency response curves of the LED-LED link under different optical power. With the decreasing of optical power, the total response intensity decreases but the response curve becomes flatter and can reach a larger bandwidth.

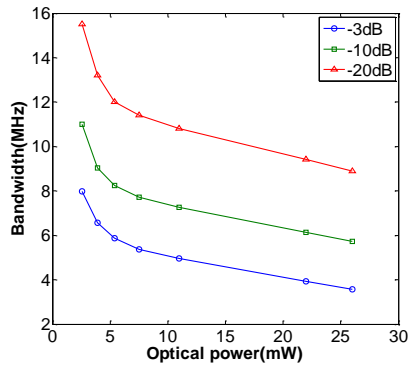

Fig. 7. Bandwidth changes with optical power for the orange-red LED to orange-red LED link.

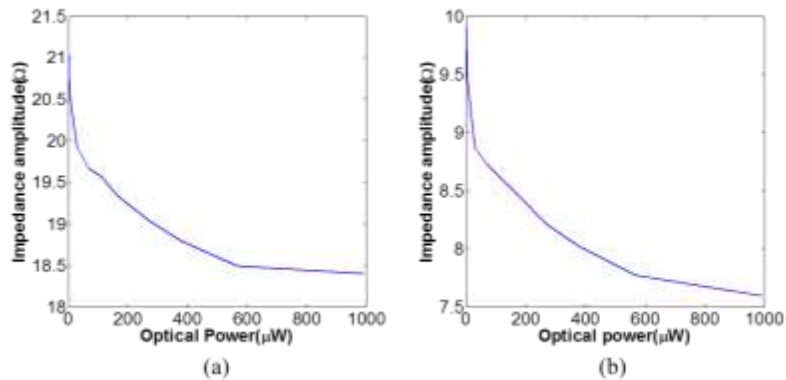

Fig. 8. Impedance changes with optical power. (a) At frequency 8MHz. (b) At frequency 10MHz.

The bandwidth variation is mainly because the AC impedance of the LED receiver changes with the optical power, as shown in Figure 8. At the low power level, the larger optical power leads to the lower impedance amplitude of the LED receiver. The influence of the AC impedance spectra of the LED receiver on the frequency response of the LED-LED link suggests that an impedance-matching TIA is necessary to post-equalize the LED-LED communication channel for increased bandwidth.

**Influence on LED-LED communication**

On one hand, the optical power reduction causes the bandwidth of the LED-LED link increased, while on the other hand it also reduces the SNR. As is known, both SNR and bandwidth affect the performance of the LED-LED VLC system. Nevertheless, some modulation schemes might be more sensitive to SNR and the others might be more sensitive to the bandwidth. Here we applied the experimental results to LED-LED communication using three modulations: OOK, 4PAM, and 8PAM. The bit rate when BER equals $1\times10^{-3}$ is used to evaluate their performances. Gray code is used in PAM to improve BER performance, for example, the Gray coed for 4PAM is shown in Table 1.

Table 1. Gray code for 4PAM

| 4PAM level | Gray code |
| --- | --- |
| -3 | 00 |
| -1 | 01 |
| 1 | 11 |
| 3 | 10 |

To study the effects of LED receiver impedance on the LED-LED communication performance study, we apply IFFT to the measured frequency response curves of the LED-LED link to obtain the time domain impulse response. A binary input sequence is input to the above channel whose output at the LED receiver is fed to a TIA. In the simulation, the parameters of the LED transmitter keep fixed, and both the photoelectric response efficiency of the LED receiver and the magnification of the TIA are assumed to be independent of the optical power. The received optical power is changed via adjusting the distance between transmitter and receiver. In this situation, the peak-to-peak voltage $V_{pp}$ of the received electrical signal is proportional to the received average optical power $P_{optical}$, as described by Eq. (9)

$$V_{pp} \propto V_{DC} \propto I_{photo} \propto P_{optical} \tag{9}$$

where $V_{DC}$ is the direct current (DC) voltage component of the received electrical signal, and $I_{photo}$ is the average output photocurrent of the LED receiver. The receiver noise variance $\sigma^2$ is fixed and shot noise is ignored in the simulation. The SNR is defined as

$$\text{SNR} = 20\log_{10}\frac{V_{PP}}{\sigma} \tag{10}$$

In an offline experiment, SNR is measured as 41dB at $P_{optical}$ of 26mW. Figure 9 shows the simulated bit rate results for fixed BER of $1\times10^{-3}$ corresponding to OOK, 4PAM and 8PAM respectively. In general, the bit rate of each modulation firstly increases and then decreases with the average optical power. When optical power is high, the channel bandwidth is small and inter-symbol interference (ISI) limits the data rate. When optical power is low, SNR is small and thus needs to lower data rate to maintain the BER performance. There exists the best point to balance the bandwidth and SNR. Because different modulations have different sensitivities to bandwidth and SNR, their trends are distinct. OOK works well in the low optical power region while higher order PAM works better in the higher optical power region. For a fixed optical power, one can choose a modulation to achieve a better performance.

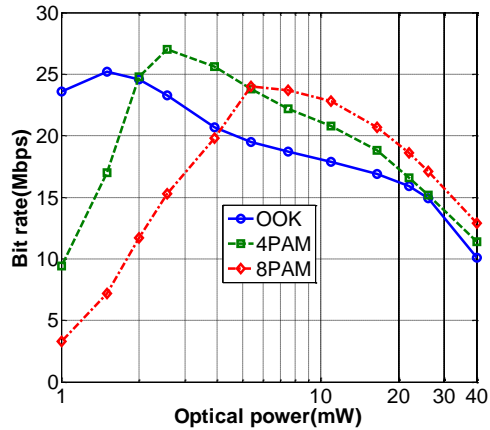

Fig. 9. Bit rate versus the average received optical power for different modulations.

## Conclusion

We experimentally demonstrate the AC impedance of the LED receiver heavily depends on the injected optical power, and provide a theoretical interpretation by an analytical model. The light-dependent impedance of the LED receiver results in the bandwidth decay of the LED-LED visible light communication link as the received optical power increases. The LED-LED communication performance of three modulations: OOK, 4PAM, and 8PAM are simulated based on the experimental frequency response curves. There exists an optimal injection power to achieve the highest data rate for a fixed BER requirement. Future research will further investigate the effects of the LED receiver impedance when other advanced modulation schemes such as OFDM and its variants are used.

## Acknowledgement

This work was supported by National Key Basic Research Program of China (Grant No. 2013CB329201), Key Program of National Natural Science Foundation of China (Grant No. 61631018), National Natural Science Foundation of China (Grant No. 61501420), Key Research Program of Frontier Sciences of CAS (Grant No. QYZDY-SSW-JSC003), Key Project in Science and Technology of Guangdong Province (Grant No. 2014B010119001), and Shenzhen Peacock Plan (No. 1108170036003286).